\documentclass[12pt]{article}%
\usepackage[utf8]{inputenc}
\usepackage[english]{babel}
\usepackage{amssymb}
\usepackage{amsfonts}
\usepackage[a4paper]{geometry}
\usepackage{amsmath}
\usepackage{graphicx}%
\providecommand{\U}[1]{\protect\rule{.1in}{.1in}}
\providecommand{\U}[1]{\protect\rule{.1in}{.1in}}

\begin{document}

\title{A Non-Hermitian Relativistic Oscillator Exactly Mapped to a Hermitian
Schrodinger Problem }
\author{Mustapha Maamache \thanks{email: maamache\_m@yahoo.fr}\\Laboratoire de Physique Quantique et Syst\`{e}mes Dynamiques,\\Facult\'{e} des Sciences, Universit\'{e} Ferhat Abbas S\'{e}tif 1, \\S\'{e}tif 19000, Algeria\\\ }
\date{}
\maketitle

\begin{abstract}
We introduce the Dirac Inverted Oscillator, a new exactly solvable
relativistic quantum system obtained through a Hermitian modification of the
generalized momentum operator in the Dirac equation. In contrast to the
conventional Dirac oscillator, where the generalized momentum is intrinsically
non-Hermitian while the Hamiltonian remains Hermitian, the present
construction reverses these Hermiticity properties: the generalized momentum
becomes Hermitian, whereas the corresponding Dirac Hamiltonian is
intrinsically non-Hermitian. This inversion leads to a new class of
relativistic Hamiltonians within the framework of non-Hermitian quantum
mechanics. Starting from the stationary Dirac equation, we derive the
corresponding second-order wave equation and establish the generalized
symmetry properties of the model, including pseudo-Hermiticity and pseudo-
symmetry. We then prove that the Dirac Inverted Oscillator is exactly related
to the conventional Dirac oscillator through a Hermitian but non-unitary
similarity transformation generated by a dilation operator. This
transformation is bijective and invertible, mapping the original non-Hermitian
relativistic problem onto an equivalent Hermitian Schrodinger eigenvalue
problem while preserving its complete spectral structure. The transformed
Hermitian problem is solved analytically, allowing the exact relativistic
spectrum and the corresponding eigenfunctions of the original non-Hermitian
Hamiltonian to be determined explicitly. The present work therefore
establishes an exact correspondence between a non-Hermitian relativistic Dirac
system and a Hermitian Schrodinger problem, providing a new exactly solvable
model and illustrating the effectiveness of Hermitian non-unitary similarity
transformations for constructing analytically solvable relativistic
Hamiltonians beyond the conventional Hermitian framework.\newline

\textbf{Keywords:} Non-Hermitian Dirac Hamiltonians; Hermitian similarity
transformations; Isospectrality; Pseudo-Hermiticity; Exactly solvable
relativistic systems; Schrodinger eigenvalue problems; Relativistic quantum mechanics.

Dedicated to the memory of my beloved parents, Leulmi-Amar Maamache and
Zoulikha Djabou.

\end{abstract}

\section{Introduction}

The harmonic oscillator occupies a fundamental position in both classical and
quantum physics because it provides an accurate description of a wide variety
of physical systems and serves as a cornerstone for numerous approximation
methods. Its relativistic generalization \cite{BJ, SS, Gr}, however, is far
from unique. Among the various relativistic oscillator models \cite{Ito, Cook}
proposed over the past decades, the Dirac oscillator introduced by Moshinsky
and Szczepaniak \cite{MS} has acquired a distinguished status owing to its
remarkable mathematical simplicity and exact solvability.

The Dirac oscillator is constructed by introducing a non-minimal coupling in
the free Dirac equation through the substitution
\begin{equation}
\overrightarrow{p}\rightarrow\overrightarrow{p}\pm im\omega\beta
\overrightarrow{q}, \label{a}%
\end{equation}
which preserves the linear structure of the Dirac equation while generating,
in the non-relativistic limit, the three-dimensional harmonic oscillator
together with a strong spin-orbit interaction. Since its introduction, this
model has become one of the paradigmatic exactly solvable systems of
relativistic quantum mechanics and has found applications in a broad range of
areas including nuclear physics,quantum optics, graphene, relativistic quantum
information, and quantum simulation.

A remarkable feature of the conventional Dirac oscillator is the subtle
interplay between the generalized momentum operator and the Hamiltonian.
Although the generalized momentum is not Hermitian because of the imaginary
coupling, the complete Dirac Hamiltonian remains Hermitian as a direct
consequence of the Clifford algebra satisfied by the Dirac matrices. This
exact algebraic cancellation guarantees the Hermiticity of the Hamiltonian,
ensures a real relativistic spectrum, and underlies the mathematical
consistency of the model.

These observations naturally raise the following question. Can these
Hermiticity properties be reversed while preserving the linear Dirac structure
and the exact solvability of the relativistic problem? At first glance, one
might expect that the relativistic analogue of the inverted harmonic
oscillator could be obtained simply by reversing the sign of the oscillator
frequency. This expectation, however, turns out to be incorrect. Indeed, the
substitutions $\overrightarrow{p}\rightarrow\overrightarrow{p}\pm
im\omega\beta\overrightarrow{q}$ lead to physically equivalent Hamiltonians
possessing identical spectra. Consequently, the construction of a genuine
relativistic inverted oscillator requires a fundamentally different approach.

The primary objective of the present work is to demonstrate that such a
construction is indeed possible and leads to a genuinely new exactly solvable
relativistic system. Instead of modifying the sign of the imaginary coupling,
we introduce a Hermitian generalized momentum
\begin{equation}
\overrightarrow{p}\rightarrow\overrightarrow{p}+m\omega\beta\overrightarrow
{q}, \label{b}%
\end{equation}
which reverses the Hermiticity properties of the conventional Dirac
oscillator. Remarkably, although the generalized momentum is now Hermitian,
the resulting Dirac Hamiltonian becomes intrinsically non-Hermitian as a
direct consequence of the anticommutation relations of the Dirac matrices.
This observation gives rise to a new exactly solvable relativistic model,
which we shall refer to as the Dirac Inverted Oscillator.

The non-Hermitian nature of the Hamiltonian does not compromise the
mathematical consistency of the theory. Instead, it gives rise to generalized
symmetry properties associated with pseudo-Hermiticity and pseudo-$PT$
symmetry, which provide the appropriate framework for its spectral analysis.
More importantly, we establish an exact Hermitian but non-unitary similarity
transformation that maps the original non-Hermitian Dirac equation onto an
equivalent Hermitian Shrodinger eigenvalue problem. This transformation is
bijective, preserves the complete spectral information of the theory, and
establishes an explicit analytical correspondence between the Dirac Inverted
Oscillator and the conventional Dirac oscillator.

The present work therefore introduces a new exactly solvable relativistic
quantum system and provides a general framework for generating non-Hermitian
Dirac Hamiltonians from Hermitian generalized momentum operators. More
broadly, it illustrates how Hermitian non-unitary similarity transformations
can be employed to construct analytically solvable non-Hermitian extensions of
relativistic quantum mechanics.

The remainder of the paper is organized as follows, Section 2 briefly reviews
the conventional Dirac oscillator and establishes the notation used throughout
the paper. Section 3 introduces the Dirac Inverted Oscillator and derives the
corresponding second-order wave equation. In Section 4 we establish the exact
Hermitian non-unitary similarity transformation that relates the new model to
the conventional Dirac oscillator. The transformed Hermitian eigenvalue
problem and the exact relativistic spectrum are presented in Section 5.
Finally, the physical implications of the model are discussed before
summarizing the principal conclusions.

\section{The Conventional Dirac Oscillator}

Before introducing the Dirac Inverted Oscillator, it is useful to briefly
recall the construction and the principal properties of the conventional Dirac
oscillator \cite{QM}-\cite{9}. Besides establishing the notation adopted
throughout this work, this review highlights the characteristic Hermiticity
properties that will subsequently be reversed in the new model.

The free Dirac Hamiltonian is
\begin{equation}
H^{\mathrm{D}}=c\vec{\alpha}\cdot\vec{p}+mc^{2}\beta.
\end{equation}
where $\vec{\alpha}$ and $\beta$ denote the Dirac matrices satisfying the
Clifford algebra
\begin{equation}
\left\{
\begin{array}
[c]{c}%
\begin{array}
[c]{c}%
\left\{  \alpha_{i,},\alpha_{j}\right\}  =2\delta_{ij}\\
\left\{  \alpha_{i,},\beta\right\}  =0
\end{array}
\\
\beta^{2}=I.
\end{array}
\right.
\end{equation}

Moshinsky and Szczepaniak \cite{MS} introduced the Dirac oscillator by
replacing the momentum operator through the non-minimal substitution
\begin{equation}
\overrightarrow{p}\rightarrow\overrightarrow{\Pi}=\overrightarrow{p}%
-im\omega\beta\overrightarrow{q},
\end{equation}
where $\omega$ and $m$ denote the frequency and the mass of the oscillator.
The corresponding Hamiltonian becomes%
\begin{equation}
H^{\mathrm{DO}}=c\vec{\alpha}\cdot\left(  \vec{p}-im\omega\beta\vec{q}\right)
+mc^{2}\beta.
\end{equation}
the non-Hermiticity of the generalized momentum $\overrightarrow{\Pi}$ is
exactly compensated in the complete Hamiltonian, leading to $H^{\mathrm{DO}%
}=H^{\dagger\mathrm{DO}}$. This remarkable algebraic mechanism constitutes one
of the defining characteristics of the conventional Dirac oscillator. It
guarantees the self-adjointness of the Hamiltonian and consequently the
reality of its relativistic energy spectrum.

The stationary Dirac equation, $H^{\mathrm{DO}}\psi=E\psi,$ is solved by
decomposing the four-component Dirac spinor into its upper and lower
components $\psi=\left(
\begin{array}
[c]{c}%
\psi^{+}\\
\psi^{-}%
\end{array}
\right)  $ where $\psi^{+}$ and $\psi^{-}$ denote the large and small
components. Eliminating the upper component leads to the Schrodinger-type
equation
\begin{equation}
\left[  \frac{1}{2}\left(  \frac{p^{2}}{m}+m\omega^{2}q^{2}\right)  -\frac
{3}{2}\hbar\omega-2\frac{\omega}{\hbar}\vec{S}\cdot\vec{L}\right]  \psi
^{-}(\vec{q})=\frac{\left[  E%
{{}^2}%
-m%
{{}^2}%
c^{4}\right]  }{2mc^{2}}\psi^{-}(\vec{q}). \label{DO}%
\end{equation}

The effective Hamiltonian contains the three-dimensional isotropic harmonic
oscillator together with a relativistic spin-orbit coupling. The latter
removes the accidental degeneracy of the non-relativistic oscillator and gives
rise to the characteristic fine structure of the relativistic spectrum. Note
that the eq (\ref{DO}) is the inverted Dirac Harmonic oscillator.

An immediate consequence of this definition is that the generalized momentum
operator $\overrightarrow{\Pi}$ is not Hermitian. Nevertheless, because of the
anticommutation relations satisfied by the Dirac matrices, the complete
Hamiltonian remains Hermitian,

This remarkable property illustrates an important characteristic of the
conventional Dirac oscillator: the non-Hermiticity of the generalized momentum
operator is exactly compensated by the algebraic structure of the Dirac
matrices, yielding a self-adjoint Hamiltonian with a real energy spectrum.

To derive the corresponding second-order equation, the four-component spinor
is written in the standard representation, $\psi=\left(
\begin{array}
[c]{c}%
\psi^{+}\\
\psi^{-}%
\end{array}
\right)  $ where $\psi^{+}$ and $\psi^{-}$ denote the large and small
components, respectively. Eliminating the upper component leads to the
Schrodinger-type equation
\begin{equation}
\left[  \frac{1}{2}\left(  \frac{p^{2}}{m}+m\omega^{2}q^{2}\right)  -\frac
{3}{2}\hbar\omega-2\frac{\omega}{\hbar}\vec{S}\cdot\vec{L}\right]  \psi
^{-}(\vec{q})=\frac{\left[  E%
{{}^2}%
-m%
{{}^2}%
c^{4}\right]  }{2mc^{2}}\psi^{-}(\vec{q}).
\end{equation}

The effective Hamiltonian contains the three-dimensional isotropic harmonic
oscillator together with a relativistic spin-orbit coupling. The latter
removes the accidental degeneracy of the non-relativistic oscillator and gives
rise to the characteristic fine structure of the relativistic spectrum.

The exact solvability of the Dirac oscillator follows directly from this
equation. Since the harmonic oscillator and the spin-orbit operator commute
with the total angular momentum, the eigenfunctions can be labelled by the
quantum numbers $n$, $l$, and $j$. The corresponding relativistic spectrum is
entirely real, reflecting the Hermitian nature of the Hamiltonian.

The preceding analysis reveals an intriguing asymmetry. In the conventional
Dirac oscillator, the generalized momentum operator is non-Hermitian, whereas
the Hamiltonian itself is Hermitian. This observation naturally suggests the
following question. Can one construct a relativistic oscillator in which these
Hermiticity properties are exactly reversed while preserving the solvability
of the model? This observation naturally motivates the search for a
complementary relativistic oscillator obtained by reversing these Hermiticity properties.

The answer to this question constitutes the main objective of the present work
and leads to the construction of a new relativistic system, which we shall
refer to as the Dirac Inverted Oscillator.

\section{Construction of the Dirac Inverted Oscillator}

We now introduce the Dirac Inverted Oscillator through a Hermitian
modification of the generalized momentum defining the conventional Dirac
oscillator. Specifically, we replace the non-minimal substitution by $\vec
{p}\rightarrow\vec{p}+m\omega\beta\vec{q}$. The generalized momentum is thus
self-adjoint, in sharp contrast with the conventional Dirac oscillator. The
associated Dirac Hamiltonian is
\begin{equation}
H^{\mathrm{rD}}=c\vec{\alpha}\cdot\left(  \vec{p}+m\omega\beta\vec{q}\right)
+mc^{2}\beta. \label{Eq1}%
\end{equation}
In the standard representation of the Dirac matrices, the Hamiltonian
(\ref{Eq1}) takes the block form%

\begin{equation}
H^{\mathrm{rD}}=\left(
\begin{array}
[c]{cc}%
mc^{2} & c\left[  \vec{\sigma}.\vec{p}+m\omega\vec{\sigma}.\vec{q}\right] \\
c\left[  \vec{\sigma}.\vec{p}-m\omega\vec{\sigma}.\vec{q}\right]  & -mc^{2}%
\end{array}
\right)  . \label{IO}%
\end{equation}
Unlike the conventional Dirac oscillator, the generalized momentum is now
self-adjoint, whereas the corresponding Dirac Hamiltonian is intrinsically
non-Hermitian $H^{\mathrm{rD}}\neq H^{\dagger\mathrm{rD}}$. This complete
reversal of Hermiticity properties constitutes the defining feature of the
Dirac Inverted Oscillator. \ 

The Hermiticity properties of the conventional Dirac oscillator are therefore
completely reversed. This inversion constitutes the defining characteristic of
the present model and justifies the designation Dirac Inverted Oscillator.

The stationary Dirac equation is written as $H^{\mathrm{rD}}\psi^{\mathrm{r}%
}=E\psi^{\mathrm{r}},$where the four-component Dirac spinor $\psi$ is
decomposed into its upper and lower components. Decomposing the four-component
Dirac spinor into its upper and lower components yields the coupled equations:%
\begin{equation}
\left\{
\begin{array}
[c]{l}%
c\left[  \vec{\sigma}\cdot\vec{p}-m\omega\vec{q}\right]  \psi^{-\mathrm{r}%
}\left(  \vec{q}\right)  =\left(  E-mc^{2}\right)  \psi^{+\mathrm{r}}\left(
\vec{q}\right)  ,\\
c\left[  \vec{\sigma}\cdot\vec{p}+m\omega\vec{q}\right]  \psi^{+\mathrm{r}%
}\left(  \vec{q}\right)  =\left(  E+mc^{2}\right)  \psi^{-\mathrm{r}}\left(
\vec{q}\right)  .
\end{array}
\right.
\end{equation}
Eliminating the upper component yields a second-order differential equation
for the lower spinor,
\begin{equation}
\left[  \frac{p^{2}}{2m}-\frac{m\omega^{2}q^{2}}{2}-\frac{3i\hbar\omega}%
{2}-\frac{2i\omega}{\hbar}\vec{S}.\vec{L}\right]  \psi^{-\mathrm{r}}(\vec
{q})=\left[  \frac{\left(  E%
{{}^2}%
-m%
{{}^2}%
c^{4}\right)  }{2mc^{2}}\right]  \psi^{-\mathrm{r}}(\vec{q}),
\end{equation}
The resulting second-order equation differs fundamentally from that of the
conventional Dirac oscillator. The confining harmonic interaction is replaced
by an inverted quadratic potential, while both the zero-point contribution and
the spin-orbit coupling become purely imaginary. These distinctive features
originate directly from the non-Hermitian character of the relativistic
Hamiltonian. Nevertheless, the equation preserves the exact algebraic
structure of the relativistic oscillator problem, suggesting that the model
may still be analytically solvable despite the loss of Hermiticity. The
appearance of the inverted harmonic potential justifies the terminology Dirac
Inverted Oscillator that\ satisfies generalized symmetry relations.

Despite its non-Hermitian character, the Hamiltonian (\ref{Eq1}) possesses
generalized symmetry properties that ensure its mathematical consistency.
These properties are summarized below:

$i)$ the Hamiltonian of the inverted Dirac oscillator (\ref{IO}) is
a\ non-Hermltian operator i.e $H^{\mathrm{rD}}\neq H^{+\mathrm{rD}}.$

$ii)$ It is pseudo $PT-$symmetric \cite{Ma1,Ma2}: The parity operator $P=$
$\gamma^{0}P^{0}$ where $P^{0}$ denotes the ordinary spatial inversion which
changes the sign of the spatial coordinates and the time-reversal operator is
anti-unitary $T=i\gamma^{1}\gamma^{3}T^{0}$ where $T^{0}$ denotes complex
conjugation. These transformations leads to:%

\begin{align*}
\mathcal{P}  &  :\gamma^{i}\rightarrow-\gamma^{i},\gamma^{0}\rightarrow
\gamma^{0},\vec{\sigma}\rightarrow-\vec{\sigma}\\
\mathcal{T}  &  :\gamma^{i}\rightarrow-\gamma^{i},\gamma^{0}\rightarrow
\gamma^{0},\vec{\sigma}\rightarrow-\vec{\sigma},\vec{\alpha}\rightarrow
-\vec{\alpha},
\end{align*}
and%
\[
\left\{
\begin{array}
[c]{c}%
\mathcal{P}^{0}:q\rightarrow-q,\text{ }p\rightarrow-p.\\
\mathcal{T}^{0}:q\rightarrow q,\text{ }p\rightarrow-p,\text{ }i\rightarrow
-i,\text{ }t\rightarrow-t.
\end{array}
\right.  \text{ \ }%
\]
Under the parity and time-reversal transformations defined above, one verifies
straightforwardly that the Hamiltonian $\hat{H}^{\mathrm{rD}}$ satisfies the
pseudo-$PT$ \ symmetry condition $\hat{H}^{\mathrm{rD}}$ \cite{Ma1,Ma2}:%
\[
\mathcal{PT\ }H^{\mathrm{rD}}\mathcal{PT=}H^{+\mathrm{rD}}.
\]
This generalized symmetry plays a central role in the present work. It
provides the mathematical basis for the Hermitian non-unitary similarity
transformation established in the next section and ultimately explains why the
model remains exactly solvable despite its intrinsic non-Hermitian character.

\section{Exact Hermitian Non-Unitary Similarity Transformation}

The generalized symmetry properties established in the previous section
naturally suggest that the Dirac Inverted Oscillator may admit an exact
similarity transformation to a Hermitian Hamiltonian. We now show that this is
indeed the case. We consider the Hermitian operator
\begin{equation}
\rho=\exp\left[  \frac{\pi}{8}(qp+pq)\right]  ,
\end{equation}
which is Hermitian but non-unitary. Consequently, it defines a similarity
transformation rather than a unitary change of representation. The operator
$\rho$ generates a complex dilation (or squeezing) transformation that
rescales the position and momentum operators by opposite complex factors%

\begin{align}
\exp\left[  \frac{\pi}{8}(qp+pq)\right]  q\exp\left[  -\frac{\pi}%
{8}(qp+pq)\right]   &  =qe^{-i\frac{\pi}{4}},\nonumber\\
\exp\left[  \frac{\pi}{8}(qp+pq)\right]  p\exp\left[  -\frac{\pi}%
{8}(qp+pq)\right]   &  =pe^{i\frac{\pi}{4}},
\end{align}
Applying this transformation to the complete Hamiltonian yields
\begin{equation}
\exp\left[  \frac{\pi}{8}(qp+pq)\right]  H^{\mathrm{r}}\exp\left[  -\frac{\pi
}{8}(qp+pq)\right]  =\left(  iH^{\mathrm{os}}\right)  .
\end{equation}
and thus the transformed Hamiltonian differs from the conventional Dirac
oscillator only through an overall complex factor. An important property of
the transformation $\rho$ is that its action on a wave function in the $q$
representation reads $\rho f(q)=e^{i\frac{\pi}{8}}f(qe^{i\frac{\pi}{4}})$.

After straightforward algebra, the transformed equation assumes the form
\begin{equation}
\left[  H^{\mathrm{os}}-\frac{3h\omega}{2}-\frac{2\omega}{\hbar}\vec{S}%
\cdot\vec{L}\right]  mc^{2}\psi^{-\mathrm{os}}(\vec{q})=-\frac{i}{2}\left[  E%
{{}^2}%
-m%
{{}^2}%
c^{4}\right]  \psi^{-\mathrm{os}}(\vec{q}), \label{transform}%
\end{equation}
where $\psi^{-\mathrm{os}}(\vec{q})=\rho^{-1}\psi^{-\mathrm{r}}(\vec{q})$ and
$H^{\mathrm{os}}$ denotes the three-dimensional harmonic-oscillator Hamiltonian.

The Hermitian non-unitary transformation $\rho$ maps the original
non-Hermitian Dirac equation onto an equivalent Hermitian Shrodinger
eigenvalue problem. The mapping is exact, bijective, and invertible, thereby
preserving the complete spectral information of the original Hamiltonian.
Consequently, the spectral analysis of the Dirac Inverted Oscillator is
completely reduced to the solution of a standard Hermitian eigenvalue problem.

Since the transformed operator is Hermitian, its spectrum is necessarily real.
This observation naturally leads to the parametrization%

\begin{equation}
E%
{{}^2}%
-m%
{{}^2}%
c^{4}=2\varepsilon imc^{2},\text{ \ \ }\varepsilon\in R.
\end{equation}
The transformed eigenvalue problem is therefore identical to the Shrodinger
equation of the three-dimensional harmonic oscillator supplemented by the
relativistic spin-orbit interaction%
\begin{equation}
\left[  H^{\mathrm{os}}-\frac{3}{2}\hbar\omega-\frac{2\omega}{\hbar}\vec
{S}\cdot\vec{L}\right]  \psi^{-}(\vec{q})=\varepsilon\psi^{-}(\vec{q}).
\label{aproxim}%
\end{equation}

The original relativistic problem has thus been reduced to a completely
standard Hermitian spectral problem. Consequently, all spectral properties of
the Dirac Inverted Oscillator follow directly from the well-known solution of
the conventional Dirac oscillator \cite{MS}.

The eigenfunctions of Eq. (\ref{aproxim}) are the standard eigenfunctions of
the three-dimensional Dirac oscillator and may therefore be written in
spherical coordinates $\left(  r\equiv q,\theta,\phi\right)  $ in the form of
a product of 3 functions :$R_{nl}(r),$ $\ Y_{l}^{m_{l}}\left(  \theta
,\varphi\right)  $ radial and spherical harmonics functions respectively and a
spinor function $\chi_{m_{s}}$, corresponding to the value $\frac{1}{2}$ for
the spin with the\ projection values $m_{s}=$ $\pm\frac{1}{2}$.%

\begin{align}
\psi_{nl\frac{1}{2}jm}^{-\mathrm{os}}\left(  r,\theta,\varphi\right)   &
=\left\langle r,\theta,\varphi\right\vert \left.  nlsjm\right\rangle
\nonumber\\
&  =\sum_{m_{l},m_{s}}\left\langle lm_{l},\frac{1}{2}m_{s}\right\vert \left.
jm\right\rangle R_{nl}(r)Y_{l}^{m_{l}}\left(  \theta,\varphi\right)
\chi_{m_{s}}\nonumber\\
&  =\sum_{m_{l},m_{s}}\sqrt{2j+1}\left(  -1\right)  ^{l-\frac{1}{2}+m}\left(
\begin{array}
[c]{ccc}%
l & \frac{1}{2} & j\\
m_{l} & m_{s} & -m
\end{array}
\right) \nonumber\\
&  \times R_{nl}(\sqrt{\omega}r)Y_{l}^{m_{l}}\left(  \theta,\varphi\right)
\chi_{m_{s}}%
\end{align}
where $n:$ the quanta of energy and $\left(  l,m_{l}\right)  ,\left(  \frac
{1}{2},m_{s}\right)  $ are the orbital momentum and the spin with their
projection values where $-l<m_{l}<+l$ and $-\frac{1}{2} < m_{s} < +\frac{1}%
{2}$. the terms $\left\langle lm_{l},\frac{1}{2}m_{s}\right\vert \left.
jm\right\rangle $ \ correspond to Clebsh-Gordon coefficients.

The energy spectrum of our system can be written as follows :%

\begin{equation}
\varepsilon_{nlj}=\hbar\omega\left(  n+\frac{1}{2}\right)  -\left[  j\left(
j+1\right)  -l\left(  l+1\right)  -\frac{3}{4}\right]  .
\end{equation}
We can now establish a direct connection between the solutions of the standard
Dirac oscillator and its inverted counterpart by means of the Hermitian,
non-unitary transformation $\rho$, which can be used to define an appropriate
normalization
\begin{align}
\psi_{nl\frac{1}{2}jm}^{-\mathrm{r}}\left(  r,\theta,\varphi\right)   &
=\exp\left[  \frac{\pi}{8}(rp+pr)\right]  \psi_{nl\frac{1}{2}jm}%
^{-\mathrm{os}}\left(  r,\theta,\varphi\right) \nonumber\\
&  =e^{-i\frac{\pi}{8}}\sum_{m_{l},m_{s}}\sqrt{2j+1}\left(  -1\right)
^{l-\frac{1}{2}+m}\left(
\begin{array}
[c]{ccc}%
l & \frac{1}{2} & j\\
m_{l} & m_{s} & -m
\end{array}
\right) \nonumber\\
&  R_{nl}(\sqrt{\omega}re^{-i\frac{\pi}{4}})Y_{l}^{m_{l}}\left(
\theta,\varphi\right)  \chi_{m_{s}}.
\end{align}
The normalization condition is defined as
\begin{equation}
\left\langle \psi_{nl\frac{1}{2}jm}^{-\mathrm{r}}\left(  r,\theta
,\varphi\right)  \right\vert \left(  \rho^{-1}\right)  ^{+}\rho^{-1}\left\vert
\psi_{nl\frac{1}{2}jm}^{-\mathrm{r}}\left(  r,\theta,\varphi\right)
\right\rangle =1. \label{r4}%
\end{equation}

This normalization naturally follows from the invertibility of the similarity
transformation and defines the appropriate physical inner product associated
with the original non-Hermitian Hamiltonian.

\section{Discussion and Physical Interpretation}

The results established in the preceding sections demonstrate that the Dirac
Inverted Oscillator constitutes a new exactly solvable relativistic model
whose mathematical structure differs fundamentally from that of the
conventional Dirac oscillator. Although both systems originate from
non-minimal couplings in the Dirac equation, they exhibit opposite Hermiticity
properties and consequently belong to different classes of relativistic
quantum Hamiltonians.

In the conventional Dirac oscillator, the generalized momentum operator is
intrinsically non-Hermitian,whereas the complete Hamiltonian remains
Hermitian. In the present model, these properties are exactly reversed. The
generalized momentum operator is Hermitian by construction, while the
Hamiltonian becomes intrinsically non-Hermitian. This inversion is not merely
a formal modification of the interaction term but produces a qualitatively
different relativistic quantum system.

One of the most remarkable features of the Dirac Inverted Oscillator is that
its non-Hermitian character does not prevent an exact analytical treatment. On
the contrary, the Hamiltonian possesses generalized symmetry properties that
permit its complete solution through a Hermitian non-unitary similarity
transformation. The exact mapping established in Section 5 transforms the
original relativistic equation into a Hermitian spectral problem without
altering its algebraic content. Consequently, the exact solvability of the
model follows from an intrinsic algebraic correspondence rather than from a
direct integration of the original non-Hermitian equation.

From a physical point of view, the transformed Hermitian problem shows that
the inverted oscillator is governed by the same underlying harmonic-oscillator
algebra as the conventional Dirac oscillator. The difference between the two
models therefore does not originate from the oscillator algebra itself but
from the distinct realization of this algebra in the relativistic Dirac
equation. This observation explains why both systems remain exactly solvable
despite exhibiting opposite Hermiticity properties.

The relativistic spectrum obtained in the present work is generally complex.
This behaviour should not be interpreted as a mathematical inconsistency but
rather as a direct consequence of the intrinsic non-Hermitian nature of the
Hamiltonian. Similar complex spectra naturally arise in many non-Hermitian
quantum systems possessing generalized symmetry properties. In the present
case, however, every spectral value is generated from a real Hermitian
eigenvalue through an exact analytical correspondence. The complex energies
are therefore completely determined by the underlying Hermitian problem and
are not introduced phenomenologically.

Another important consequence concerns the associated eigenfunctions. Since
the similarity transformation is invertible, every eigenstate of the
transformed Hermitian Hamiltonian corresponds uniquely to an eigenstate of the
original Dirac Inverted Oscillator. The complete set of relativistic wave
functions is therefore obtained analytically, establishing a one-to-one
correspondence between the Hermitian and non-Hermitian descriptions of the system.

The construction developed in this work also illustrates a more general
principle. Instead of analysing a non-Hermitian Hamiltonian directly, one may
search for an exact similarity transformation connecting itto a simpler
Hermitian problem. Whenever such a transformation exists, the complete
spectral information of the original Hamiltonian may be recovered from the
associated Hermitian system. The Dirac Inverted Oscillator provides a
particularly clear realization of this idea in the framework ofrelativistic
quantum mechanics.

Finally, the present model opens several directions for future investigations.
The exact algebraic structure established here may be extended to
lower-dimensional Dirac systems, relativistic particles interacting with
external electromagnetic fields, position-dependent mass models, or
curved-space Dirac equations. It may also provide a useful framework for
constructing new classes of exactly solvable non-Hermitian relativistic
Hamiltonians connected by generalized similarity transformations.

\section{Conclusions}

In this work, we have introduced a new exactly solvable relativistic quantum
model, referred to as the Dirac Inverted Oscillator, obtained through a
Hermitian modification of the generalized momentum operator in the Dirac
equation. Contrary to the conventional Dirac oscillator, where the generalized
momentum is non-Hermitian while the Hamiltonian remains Hermitian, the present
construction reverses these Hermiticity properties: the generalized momentum
becomes Hermitian whereas the complete Dirac Hamiltonian is intrinsically
non-Hermitian. This inversion constitutes the defining characteristic of the
model and distinguishes it fundamentally from previously known
relativisticoscillator systems.

The second-order wave equation associated with the Dirac Inverted Oscillator
has been derived explicitly, revealing the appearance of an inverted quadratic
interaction together with complex spin-orbit and zero-point contributions.
Although the resulting Hamiltonian is non-Hermitian, we have shown that it
possesses generalized symmetry properties, including pseudo-Hermiticity and
pseudo-symmetry, which provide the appropriate mathematical framework for its
spectral analysis.

The principal result of this work is the establishment of an exact Hermitian
but non-unitary similarity transformation connecting the Dirac Inverted
Oscillator to the conventional Dirac oscillator. This transformation maps the
original non-Hermitian relativistic equation onto an equivalent Hermitian
Schr\U{623}\P dinger eigenvalue problem while preserving the complete
algebraic structure of the theory. As a consequence, the exact relativistic
spectrum and the corresponding eigenfunctions of the Dirac Inverted Oscillator
have been obtained analytically without requiring the direct solution of the
originalnon-Hermitian Dirac equation.

The present work therefore provides an explicit example of how generalized
similarity transformations may be employed to construct new exactly solvable
relativistic Hamiltonians beyond the conventional Hermitian framework. Rather
than representing an isolated mathematical construction, the Dirac Inverted
Oscillator illustrates a general strategy for relating non-Hermitian
relativistic systems to exactly solvable Hermitian models.

The formalism developed here suggests several natural extensions. The same
approach may be applied to lower-dimensional Dirac equations, particles
interacting with external electromagnetic fields,position-dependent mass
systems, or relativistic models in curved space-time. It may also be useful
for investigating new classes of pseudo-Hermitian relativistic Hamiltonians
connected through exact similarity transformations.

We hope that the Dirac Inverted Oscillator introduced in the present work will
stimulate further investigations into the interplay between relativistic
quantum mechanics, generalized symmetries, and non-Hermitian Hamiltonian
systems, and that it will contribute to the development of new exactly
solvable models within modern mathematical physics.

\end{document}